\documentclass[twocolumn,showpacs,preprintnumbers,prl,amsmath,amssymb,superscriptaddress]{revtex4}
\usepackage{graphicx}
\usepackage{endnotes}
\usepackage{bm}
\usepackage{epsfig}

\begin{document}

%%%%%%%%%%%%%%%%%%%%% TITLE %%%%%%%%%%%%%%%%%%%%

\title{Chemical Pressure and Physical Pressure in BaFe$_2$(As$_{1-x}$P$_x$)$_2$}

%%%%%%%%%%%%%%%%%%%% AUTHORS %%%%%%%%%%%%%%%%%%

\author{Lina~E.~Klintberg}
\affiliation{Cavendish Laboratory, University of Cambridge, J.J. Thomson Avenue, Cambridge CB3 0HE, United Kingdom}

\author{Swee~K.~Goh}
\email{skg27@cam.ac.uk}
\affiliation{Cavendish Laboratory, University of Cambridge, J.J. Thomson Avenue, Cambridge CB3 0HE, United Kingdom}

\author{Shigeru~Kasahara}
\affiliation{Research Center for Low Temperature and Materials Sciences, Kyoto University, Kyoto 606-8502, Japan}

\author{Yusuke~Nakai}
\author{Kenji~Ishida}
\affiliation{Department of Physics, Graduate School of Science, Kyoto University, Kyoto 606-8502, Japan}
\affiliation{TRIP, JST, Sanban-cho Building, 5, Sanban-cho, Chiyoda, Tokyo 102-0075, Japan}

\author{Michael~Sutherland}
\affiliation{Cavendish Laboratory, University of Cambridge, J.J. Thomson Avenue, Cambridge CB3 0HE, United Kingdom}

\author{Takasada~Shibauchi}
\author{Yuji~Matsuda}
\affiliation{Department of Physics, Graduate School of Science, Kyoto University, Kyoto 606-8502, Japan}

\author{Takahito~Terashima}
\affiliation{Research Center for Low Temperature and Materials Sciences, Kyoto University, Kyoto 606-8502, Japan}

\date{October 21, 2010}

\begin{abstract}
Measurements of the superconducting transition temperature, $T_c$, under hydrostatic pressure via bulk AC susceptibility were carried out on several concentrations of phosphorous substitution in BaFe$_2$(As$_{1-x}$P$_x$)$_2$.  The pressure dependence of unsubstituted BaFe$_2$As$_2$, phosphorous concentration dependence of BaFe$_2$(As$_{1-x}$P$_x$)$_2$, as well as the pressure dependence of BaFe$_2$(As$_{1-x}$P$_x$)$_2$ all point towards an identical maximum $T_c$ of 31 K. This demonstrates that phosphorous substitution and physical pressure result in similar superconducting phase diagrams, and that phosphorous substitution does not induce substantial impurity scattering.
\end{abstract}


\maketitle

The discovery of pnictide superconductors \cite{Kamihara} provoked a comprehensive overhaul of the understanding of the structural and magnetic environments in which unconventional superconductivity may arise. In BaFe$_2$As$_2$, superconductivity can be induced not only by charge doping \cite{Rotterprl,Sasmal} as with cuprates, but by isoelectronic substitution\cite{Kasahara10} and pressure\cite{Alireza,Yama,Colombier} as well. In the case of BaFe$_2$(As$_{1-x}$P$_x$)$_2$, introducing smaller phosphorous atoms in the place of the arsenic atoms results in a reduction in the unit cell volume without the introduction of additional carriers into the FeAs layer, yet still enhances the superconducting transition temperature, $T_c$ \cite{Kasahara10}.

At 140~K BaFe$_2$As$_2$ concurrently undergoes an antiferromagnetic (AFM) spin density wave (SDW) transition and a structural transition\cite{Baek}. This SDW state is observed for phosphorous content $x\leq 0.27$ whereas superconductivity can be seen for $0.14\leq x\leq 0.71$ with a maximum $T_c$ of 31~K at $x\sim 0.3$ \cite{Kasahara10}.
The suppression of the SDW state is essential for the emergence of superconductivity, and this suppression is intimately linked with the iron-pnictogen distance, not the total cell volume\cite{Rotter}. This is why chemical pressure in Ba$_{1-x}$Sr$_x$Fe$_2$As$_2$ does not result in superconductivity in spite of similarities -- in both unit cell volume and cell volume reduction rate as a function of isovalent substitution -- with BaFe$_2$(As$_{1-x}$P$_x$)$_2$. This strong dependence on the Fe-Pn distance is presumably why uniaxial stress strongly suppresses the structural as well as the AFM ordering, and stabilises superconductivity in BaFe$_2$As$_2$ \cite{Yama}. Recently, AC susceptibility measurements were conducted on optimally substituted BaFe$_2$(As$_{0.65}$P$_{0.35}$)$_2$ to study the evolution of $T_c$ both as a function of pressure and applied magnetic field. Interestingly, although the electronic structure is thought to become more isotropic under pressure, the superconducting properties become more anisotropic \cite{Goh}. Another interesting feature of BaFe$_2$(As$_{1-x}$P$_x$)$_2$ is that there is evidence for line nodes in the energy gap \cite{Hashimoto,Nakaiprb} as has also been seen in KFe$_2$As$_2$ \cite{Fukazawa,Hashimoto2}.

Comparisons of chemical and physical pressure on the sister pnictide EuFe$_2$As$_2$ show that similar characteristics to BaFe$_2$As$_2$ are seen at low pressures \cite{Miclea,Terashima}. Emergence of superconductivity is seen with the suppression of a SDW state, and it is found to superconduct with the application of both physical and chemical pressure, \cite{Uhoya,Ren,Sun,Matsubayashi,Guo} so it is natural to draw parallels to these sister pnictide compounds. 

High quality BaFe$_2$(As$_{1-x}$P$_x$)$_2$ crystals enable reliable $T_c$ measurements to be tracked as a function of pressure without significant broadening. Consequently, this letter aims to study the relationship between phosphorous content, $x$, and pressure, $p$, by applying pressure to several phosphorous contents of BaFe$_2$(As$_{1-x}$P$_x$)$_2$, and tracking $T_c$ as a function of both $p$ and $x$ in different parts of the phase diagram.

The single crystals used for this study were prepared from mixtures of FeAs, Fe, P (powders) and Ba (flakes) placed in an alumina crucible, sealed in an evacuated quartz tube and kept at $1150-1200\,^{\circ}\mathrm{C}$ for 12 hours, followed by a slow cooling to $900\,^{\circ}\mathrm{C}$ at the rate of $1.5\,^{\circ}\mathrm{C}$/hr. The crystals were characterised using X-ray diffraction (XRD) using a Mo K$_\alpha$ source as well as with resistivity measurements \cite{Kasahara10,Kasahara10b}.
All the peaks evaluated by 4-circular XRD are very sharp and for instance the phosphorous content $x$ = 0.48 samples have a typical FWHM of 0.25--0.29$^\circ$ even though it is an alloy system, demonstrating the high quality of the crystal. The As:P ratio was determined from the $z$ coordinate of pnictogen atoms in the unit cell and energy-dispersive X-ray spectroscopy. The superconducting transition at high pressure was detected using a two-coil mutual inductance technique, where a 10-turn pickup coil was placed inside the gasket hole of an anvil cell and a 140-turn modulation coil was placed around the Moissanite anvil outside the pre-indented region \cite{Goh08, Alireza03}. The crystals used were typically around ($\sim$150 $\times$ 150 $\times$ 80) $\mu$m$^3$, giving a volume filling factor of about 0.3. The modulation frequency was of the order of 15 kHz, ensuring that the skin depth (based on an average metal) is around twice the size of the relevant dimensions of the sample. Therefore, the sample is well penetrated and these are bulk AC susceptibility measurements. Glycerin was used as the pressure transmitting fluid owing to its hydrostatic nature \cite{Osakabe}, and the pressure achieved was determined by Ruby fluorescence spectroscopy.

\begin{figure}[b]
\begin{center}
\includegraphics[width=88mm]{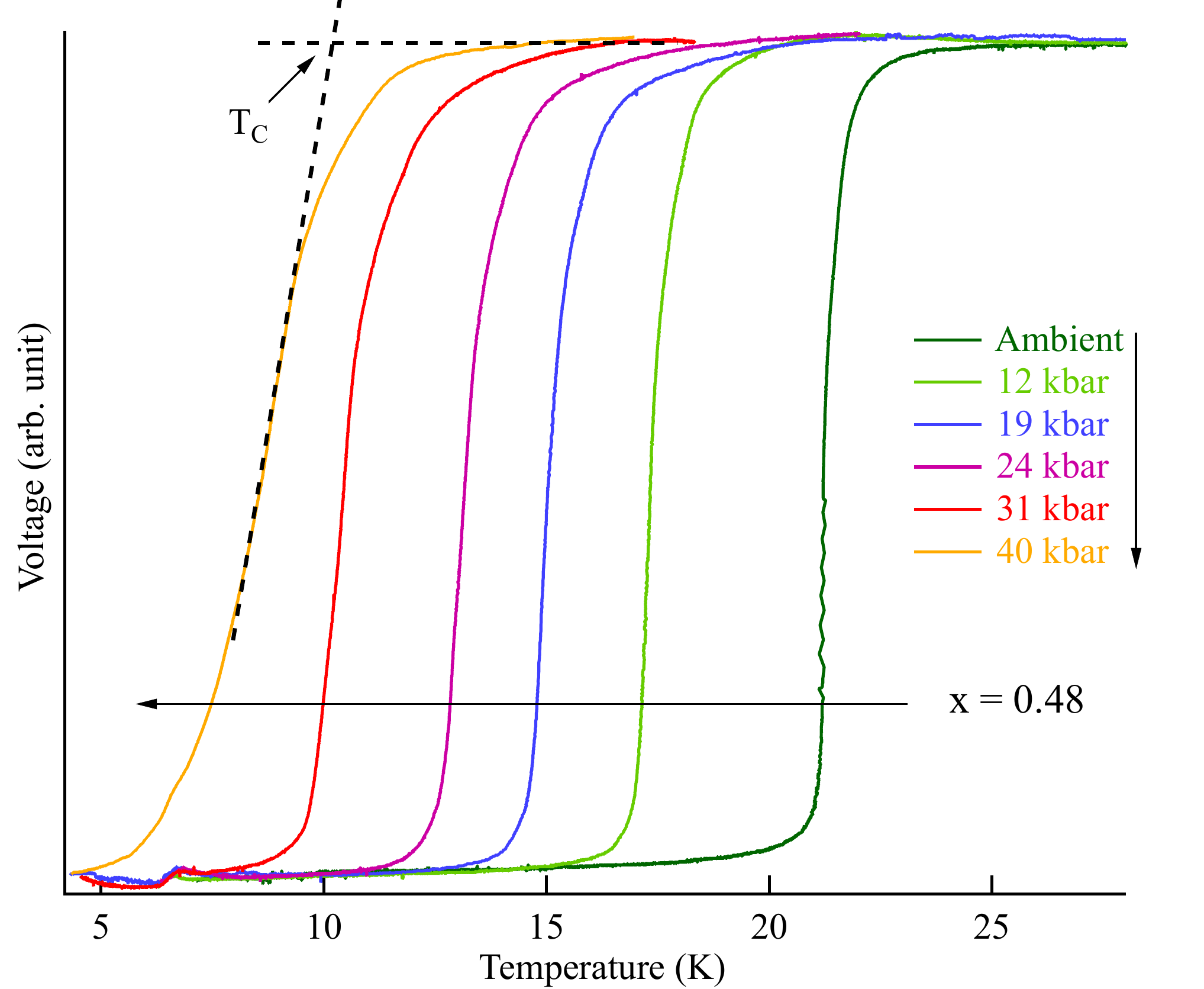}
\end{center}
\caption{(Colour online) Pressure dependence of the superconducting transition for over-substituted BaFe$_2$(As$_{1-x}$P$_x$)$_2$ at phosphorous content $x=0.48$. The small kink that can be seen at about 7 K is the superconducting transition of solder outside of the high pressure region.}
\label{f1}
\end{figure}

Figure 1 shows the normalised superconducting transitions of over-substituted BaFe$_2$(As$_{0.52}$P$_{0.48}$)$_2$ at various pressures, and Fig. 2 shows the under-substituted side at $x=0.20$ and $0.23$; these samples have structural/SDW transitions at $\sim$87/72 K and $\sim$70/58 K respectively, as measured by resistivity \cite{Kasahara10}.
Both figures clearly show the transition sharpness, which is indicative of both good hydrostaticity and high sample quality. None of the curves have been normalised by more than 1.06, as the size of the jump remains roughly constant for all measurements, arguing against large scale substitution inhomogeneities as well as phase separated antiferromagnetic islands amid superconductivity as was seen in the underdoped region of hole doped Ba$_{1-x}$K$_x$Fe$_2$As$_2$ \cite{Park}. Interestingly, the sharpest transitions all occur at maximum $T_c$ for each respective phosphorous content which is rather striking. 
Although a transition broadening is usually ascribed to pressure inhomogeneities, it cannot be argued that there is a significant broadening as a function of pressure. 
For instance, in Fig. 2, the sharpness of the 12~kbar transition is comparable to the 57~kbar transition (and similarly for the  24 and 48~kbar transitions) even though much greater broadening would be expected for the higher pressure transition if pressure inhomogeneities were significant. Furthermore, there is a clear sharpening of the transition near optimal pressures with an almost kink-like superconducting onset, suggesting that near the peak of the domes, where $dT_c/dp$ is very small, any slight sample purity inhomogeneity is undetectable and the entire sample superconducts concurrently.  On the edges of the ($T_c$,$p$)-dome where $dT_c/dp$ is significant, a small distribution of phosphorous content would cause the transition to broaden as different parts of the sample superconduct at different temperatures owing to slight inhomogeneities in $x$. In the samples measured, $x=0.20$ did not superconduct at ambient 
pressure in spite of contradictory results \cite{Kasahara10,jiang} in the literature; however, $dT_c/dp$ is very large at the onset of the superconducting dome and hence a very small discrepancy in $x$ could push the sample out of the superconducting dome completely.

\begin{figure}[t]
\begin{center}
\includegraphics[width=88mm]{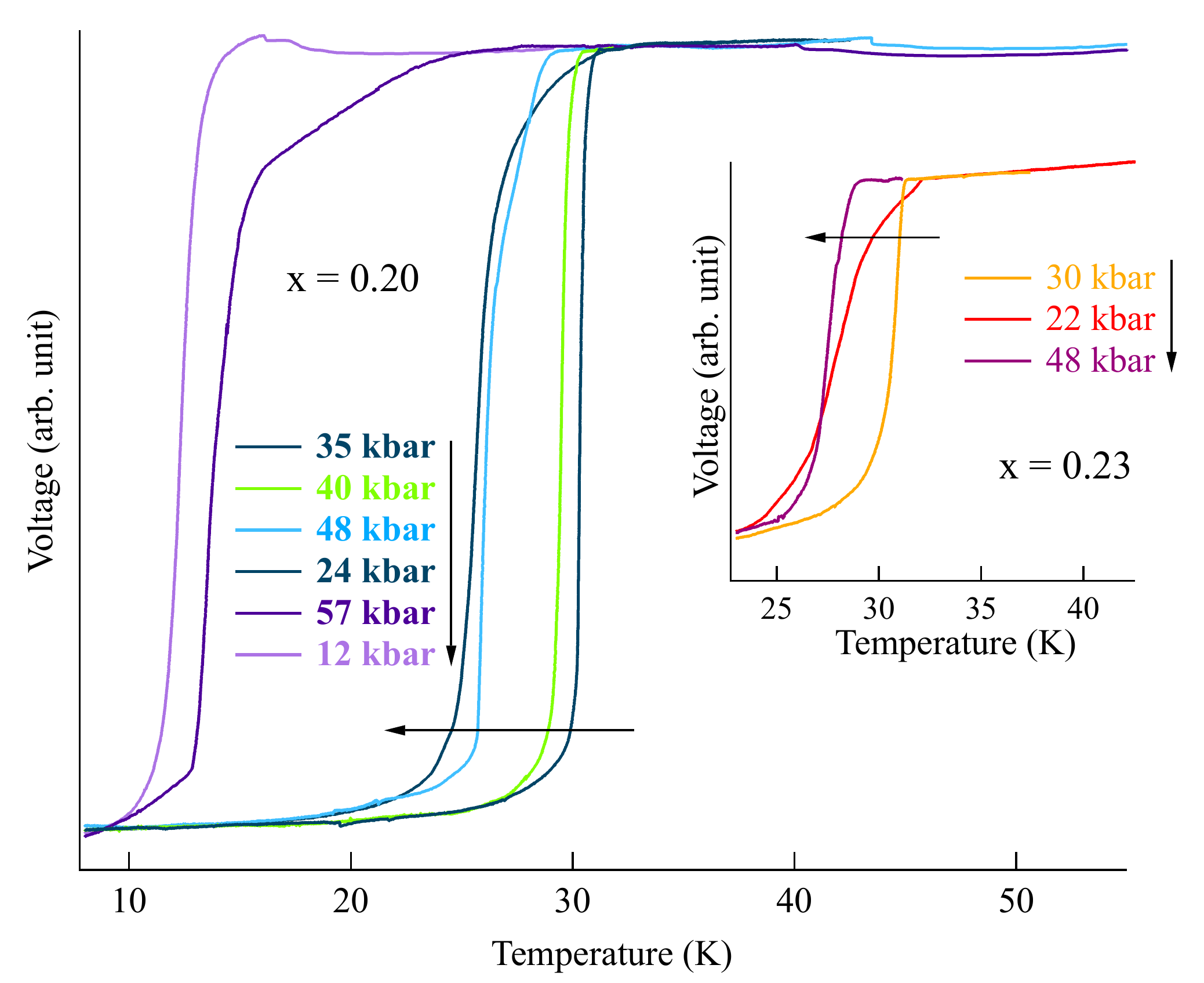}
\end{center}
\caption{(Colour online) Pressure dependence of superconducting transition for under-substituted BaFe$_2$(As$_{1-x}$P$_x$)$_2$ at phosphorous content $x=0.20$. Inset: showing only a close up of the superconducting transitions at maximum $T_c$ for $x=0.23$ as all the transitions are very similar to $x = 0.20$.}
\label{f2}
\end{figure}

Owing to the strong evidence of line nodes in the energy gap of BaFe$_2$(As$_{1-x}$P$_x$)$_2$ \cite{Fukazawa,Hashimoto2}, the aforementioned results with a common maximum $T_c$, independently of doping, indicate that phosphorous substitution does not induce substantial impurity scattering since $T_c$ should be dramatically depressed with scattering in the presence of nodes \cite{Onari}. This high sample quality is corroborated by successful quantum oscillation measurements \cite{Shishido}.

The phase diagram of $T_c$ as a function of pressure can be seen in Figs. 3 and 4, for five phosphorous contents: $x$ = 0, 0.20, 0.23, 0.35, and 0.48. The inset of Fig. 4 gives a schematic representation of the measured choices of $x$ in relation to the superconducting dome, $T_c(x)$. Figure 3 shows the un-substituted ($x = 0$) and under-substituted ($x = 0.20$ and 0.23) results, whereas Fig. 4 shows the optimally substituted ($x = 0.35$) and over-substituted ($x = 0.48$) results.
Although the maximum $T_c$ for the various phosphorous contents is a constant, applied pressure and 
chemical pressure cannot be considered interchangeable as a universal $T_c(p)$ curvature is not seen.
It is clear that as $x$ increases, the width of the dome increases from a full width of $\sim$~70 and 80~kbar  for $x = 0.20$ and 0.23 respectively, to well above 120~kbar for $x = 0.35$ and 0.48. Consequently, much higher pressure steps are required to suppress $T_c$ at higher $x$ and thus superconductivity can be considered more stable in this region. 

\begin{figure}[t]
\begin{center}
\includegraphics[width=88mm]{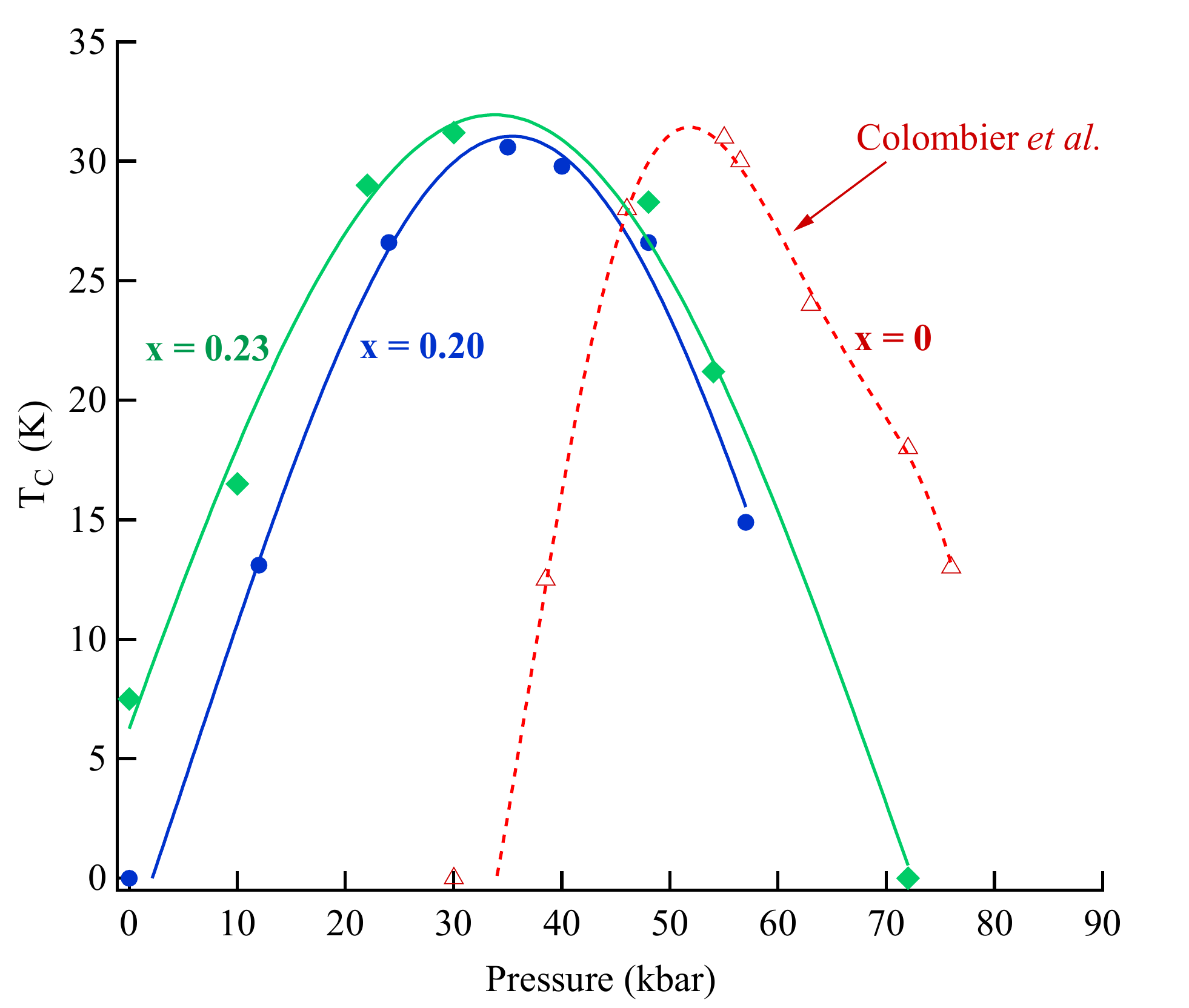}
\end{center}
\caption{(Colour online) $T_c$--pressure phase diagram of under-substituted BaFe$_2$(As$_{1-x}$P$_x$)$_2$ at phosphorous contents of $x$ = 0.20, 0.23. Data for pure BaFe$_2$As$_2$, $x$ = 0, taken from resistivity measurements \cite{Colombier} done with Fluorinert mixture 1:1 FC70:FC77 for comparison. Note that the lines are a guide to the eye.} 
\label{f3}
\end{figure}

As it has been shown that the cell volume decreases monotonically with phosphorous substitution in BaFe$_2$As$_2$ \cite{jiang}, a possible explanation for this widening of the superconducting dome as a function of $x$ is that a smaller cell volume at ambient pressure requires higher external pressure to influence the cell volume. It is possible that the suppression rate of the Fe-Pn distance as a function of pressure decreases with smaller ambient pressure cell volumes, and is no longer linear at higher levels of $x$. Detailed x-ray studies under pressure would be useful to resolve this issue.

Highly hydrostatic measurements on BaFe$_2$As$_2$ carried out using cubic-anvil cells \cite{Yama} indicate the appearance of superconductivity with much lower maximum $T_c$ of $\sim$17~K. However, an inspection of the $x$-dependence of lattice constants for BaFe$_2$(As$_{1-x}$P$_x$)$_2$ shows that the c-axis (a-axis) of BaFe$_2$P$_2$ is $\sim$4.3~\% ($\sim$3.3~\%) shorter than that of BaFe$_2$As$_2$.\cite{Kasahara10, jiang} Since the suppression of $c$ and $a$ with increasing phosphorous content $x$ follows Vegard's law, phosphorous substitution \emph{intrinsically} provides a uniaxial component throughout the entire substitution range. Therefore, while glycerin is able to provide a rather hydrostatic pressure environment, the net effect of combining chemical and physical pressure has a sizeable intrinsic uniaxial component. Consequently, it is not surprising that our present results agree better with high pressure phase diagrams of BaFe$_2$As$_2$ constructed with less hydrostatic conditions.

\begin{figure}[t]
\begin{center}
\includegraphics[width=88mm]{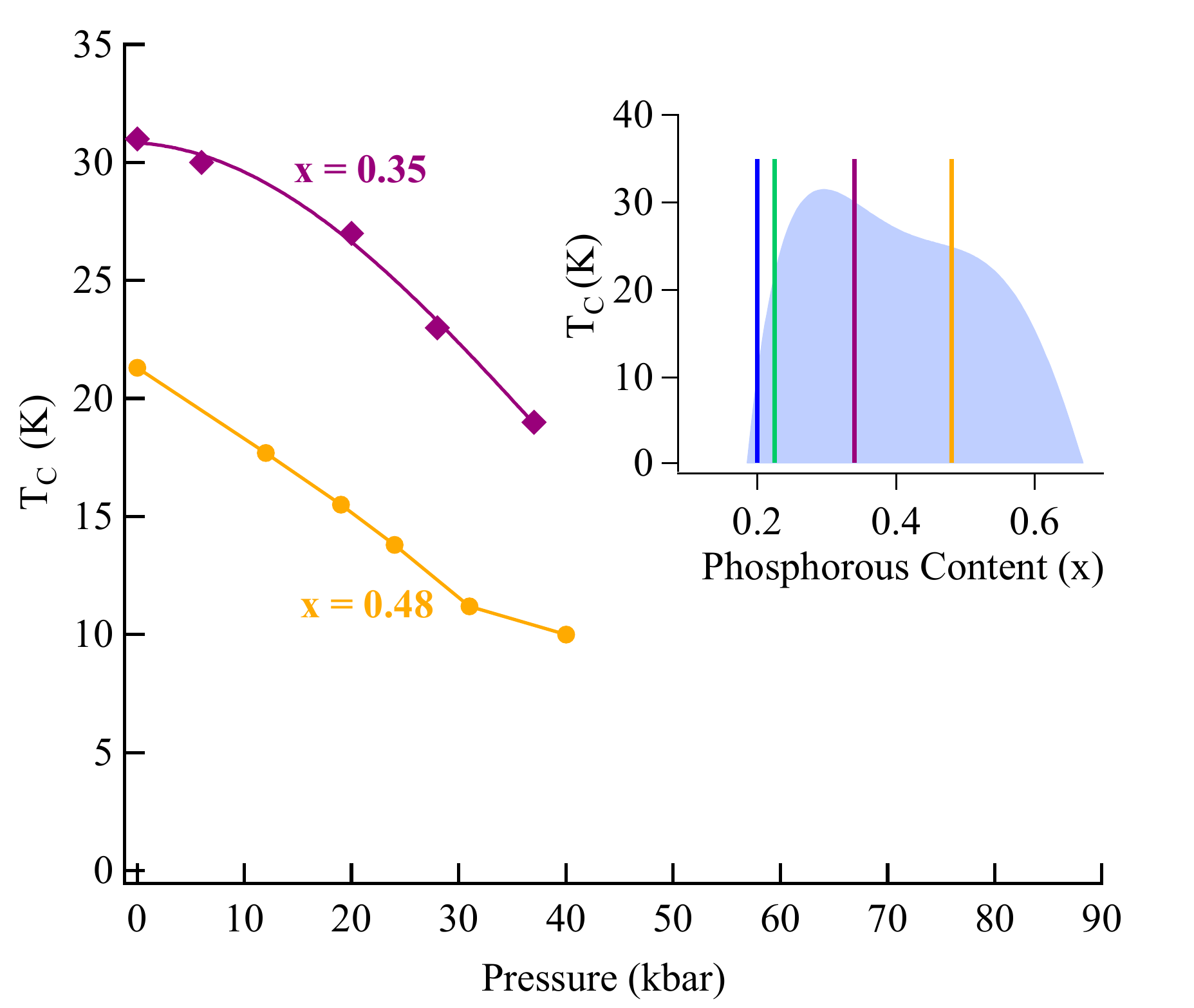}
\end{center}
\caption{(Colour online) $T_c$--pressure phase diagram of optimally and over-substituted BaFe$_2$(As$_{1-x}$P$_x$)$_2$ at $x$ = 0.35\cite{Goh} and 0.48. Inset: Schematic of the four chosen phosphorous contents in relation to the superconducting dome, with the $T_c$ vs. $x$ data taken from Ref. \cite{Nakaiprl}.} 
\label{f4}
\end{figure}

Similar studies have also been done with chemical and physical pressure on EuFe$_2$As$_2$, although 
the precise details of the phase diagram remain controversial \cite{Matsubayashi}. In some recent studies, EuFe$_2$As$_2$ is found to superconduct with $T_c$ as high as 41~K with a pressure of about 100~kbar \cite{Uhoya}. Furthermore, optimally substituted EuFe$_2$(As$_{0.7}$P$_{0.3}$)$_2$ superconducts at $\sim$26~K \cite{Ren}, which can be enhanced to 48.3~K at a pressure of about 90~kbar \cite{Guo}. These high pressure resistivity measurements need to be corroborated by detailed bulk probe measurements.

Unlike BaFe$_2$As$_2$, EuFe$_2$As$_2$ exhibits strong moments of Eu$^{2+}$ ions with AFM ordering at about 20~K in addition to the AFM Fe moment ordering. Furthermore, a valence change to a higher  Eu$^{3+}$/Eu$^{2+}$ ratio is seen under pressure in both EuFe$_2$(As$_{0.7}$P$_{0.3}$)$_2$ and the parent compound \cite{Sun}. This ratio saturates at pressures around 90~kbar: the point at which $T_c$ is found to be the largest. The divalent Eu atom is larger than the trivalent one, and therefore, not only does this valence transition transfer charge from the Eu atom to the FeAs layer, but the Eu volume change acts as additional chemical pressure. This behaviour is completely different from BaFe$_2$As$_2$. 
Consequently, although the low pressure region is quite similar for BaFe$_2$As$_2$ and EuFe$_2$As$_2$, with a SDW suppression culminating in superconductivity \cite{Miclea,Terashima} the behaviour is starkly different at higher pressures. 

It is striking that $T_c$ should increase as a function of pressure in EuFe$_2$As$_2$ if the optimal phosphorous content really has been found. One possibility is that the $T_c$ enhancement is caused by non-hydrostaticity in the pressure since BaFe$_2$As$_2$ has been shown to be very sensitive to uniaxial components \cite{Duncan,Yama}. According to theory on anvil cells using a gasket hole as the sample chamber \cite{Dunstan}, hydrostatic pressure is realised if the pressure transmitting medium has zero shear strength. Therefore, if a solid medium is used (or none at all), a distribution of pressure build up in the sample chamber, giving rise to non-hydrostaticity. 
Thermal expansion on optimally doped Ba(Fe$_{0.92}$Co$_{0.08}$)$_2$As$_2$
shows that $dT_c/dp$ is orientation dependent, and consequently that superconductivity is sensitive to uniaxial components \cite{Hardy,Budko}. $(dT_c/dp)_{tot}$ is given by the sum of $(dT_c/dp)_i$, where $i = a,b$ and $c$, and the sum of these three terms is negative near optimal doping. Therefore hydrostatic conditions give an overall depression of $T_c$. Using solid pressure media (or no medium) uniaxial pressure components are likely to build up, and dominant contributions arise from $dT_c/dp$ along one particular direction. Since $(dT_c/dp)_a$ and $(dT_c/dp)_b$ are both positive, in this scenario $T_c$ goes up. It would therefore be interesting to perform measurements on optimally substituted BaFe$_2$(As$_{1-x}$P$_{x}$)$_2$ with solid pressure media in an attempt to reproduce this $T_c$ enhancement -- as well as bulk thermodynamic measurements on EuFe$_2$(As$_{0.7}$P$_{0.3}$)$_2$ with hydrostatic pressure media -- in order to establish whether or not uniaxial components are the cause of the $T_c$ enhancement in EuFe$_2$(As$_{0.7}$P$_{0.3}$)$_2$.

In summary, the pressure dependence of un-substituted BaFe$_2$As$_2$, phosphorous concentration dependence of BaFe$_2$(As$_{1-x}$P$_x$)$_2$, as well as their pressure dependence all point towards an identical maximum $T_c$ of 31 K. This universal maximum $T_c$ demonstrates that phosphorous substitution and physical pressure result in similar superconducting phase diagrams in this compound as well as that impurity scattering does not limit $T_c$.
In spite of low pressure similarities with phosphorous substituted EuFe$_2$(As$_{1-x}$P$_x$)$_2$, the high pressure behaviour appears to be completely different.

\newpage The authors acknowledge F. M. Grosche, D. E. Khmelnitskii, P. L. Alireza and Taichi Terashima for discussions and Sam Brown for technical support. This work was supported by the EPSRC, Grants-in-Aid for Scientific Research on Innovative Areas ``Heavy Electron'' (No. 20102006) from MEXT, for the GCOE Program ÒThe Next Generation of Physics, Spun from Universality and EmergenceÓ from MEXT, and for Scientific Research from JSPS as well as Trinity College (Cambridge).

%%%%%%%%%%%%%%%%%% BIBLIOGRAPHY %%%%%%%%%%%%%%%%%%%%


\begin{thebibliography}{34}
\expandafter\ifx\csname natexlab\endcsname\relax\def\natexlab#1{#1}\fi
\expandafter\ifx\csname bibnamefont\endcsname\relax
  \def\bibnamefont#1{#1}\fi
\expandafter\ifx\csname bibfnamefont\endcsname\relax
  \def\bibfnamefont#1{#1}\fi
\expandafter\ifx\csname citenamefont\endcsname\relax
  \def\citenamefont#1{#1}\fi
\expandafter\ifx\csname url\endcsname\relax
  \def\url#1{\texttt{#1}}\fi
\expandafter\ifx\csname urlprefix\endcsname\relax\def\urlprefix{URL }\fi
\providecommand{\bibinfo}[2]{#2}
\providecommand{\eprint}[2][]{\url{#2}}




\bibitem{Kamihara}
\bibinfo{author}{Y. Kamihara, T. Watanabe, M. Hirano, H. Hosono}, 
\bibinfo{journal}{J. Am. Chem. Soc.}
\textbf{\bibinfo{volume}{130}}, \bibinfo{pages}{3296}
(\bibinfo{year}{2008}).  
  
\bibitem{Rotterprl}
\bibinfo{author}{M. Rotter, M. Tegel, and D. Johrendt}, 
\bibinfo{journal}{Phys. Rev. Lett}
\textbf{\bibinfo{volume}{101}}, \bibinfo{pages}{107006}
(\bibinfo{year}{2008}).   

\bibitem{Sasmal}
\bibinfo{author}{K. Sasmal, B. Lv, B. Lorenz, A. M. Guloy, F. Chen, Y.-Y. Xue and C.-W. Chu}, 
\bibinfo{journal}{Phys. Rev. Lett.}
\textbf{\bibinfo{volume}{101}}, \bibinfo{pages}{107007}
(\bibinfo{year}{2010}).   

\bibitem{Kasahara10}
\bibinfo{author}{S. Kasahara, T. Shibauchi, K. Hashimoto, K. Ikada, S. Tonegawa, R. Okazaki, H. Shishido, H. Ikeda, H. Takeya, K. Hirata, T. Terashima and Y. Matsuda}, 
\bibinfo{journal}{Phys. Rev. B}
\textbf{\bibinfo{volume}{81}}, \bibinfo{pages}{184519}
(\bibinfo{year}{2010}).   

\bibitem{Alireza}
\bibinfo{author}{P. L. Alireza, Y. T. C. Ko, J. Gillett, C. M. Petrone, J. M. Cole, S. E. Sebastian, and G. G. Lonzarich}, 
\bibinfo{journal}{J. Phys.: Condens. Matter}
\textbf{\bibinfo{volume}{21}}, \bibinfo{pages}{012208}
(\bibinfo{year}{2009}).  
  
\bibitem{Yama}
\bibinfo{author}{T. Yamazaki, N. Takeshita, R. Kobayashi, H. Fukazawa, Y. Kohori, K. Kihou, C.-H. Lee, H. Kito, A. Iyo, H. Eisaki}, 
\bibinfo{journal}{Phys. Rev. B}
\textbf{\bibinfo{volume}{81}}, \bibinfo{pages}{224511}
(\bibinfo{year}{2010}).  
  
\bibitem{Colombier}
\bibinfo{author}{E. Colombier, S. L. Bud'ko, N. Ni, P. C. Canfield}, 
\bibinfo{journal}{Phys. Rev. B}
\textbf{\bibinfo{volume}{79}}, \bibinfo{pages}{224518}
(\bibinfo{year}{2009}).  
 
\bibitem{Baek}
\bibinfo{author}{S.-H. Baek, T. Klimczuk, F. Ronning, E. D. Bauer, J. D. Thompson, N. J. Curro}, 
\bibinfo{journal}{Phys. Rev. B}
\textbf{\bibinfo{volume}{78}}, \bibinfo{pages}{212509}
(\bibinfo{year}{2008}).  

\bibitem{Rotter}
\bibinfo{author}{M. Rotter, C. Hieke, D. Johrendt}, 
\bibinfo{journal}{Phys. Rev. B}
\textbf{\bibinfo{volume}{82}}, \bibinfo{pages}{014513}
(\bibinfo{year}{2010}).    

\bibitem{Goh}
\bibinfo{author}{S. K. Goh, Y. Nakai, K. Ishida, L. E. Klintberg, Y. Ihara, S. Kasahara, T. Shibauchi, Y. Matsuda, and T. Terashima}, 
\bibinfo{journal}{Phys. Rev. B}
\textbf{\bibinfo{volume}{82}}, \bibinfo{pages}{094502}
(\bibinfo{year}{2010}). 

\bibitem{Hashimoto}
\bibinfo{author}{K. Hashimoto, M. Yamashita, S. Kasahara, Y. Senshu, N. Nakata, S. Tonegawa, K. Ikada, A. Serafin, A. Carrington, T. Terashima, H. Ikeda, T. Shibauchi, Y. Matsuda}, 
\bibinfo{journal}{Phys. Rev. B}
\textbf{\bibinfo{volume}{81}}, \bibinfo{pages}{220501(R)}
(\bibinfo{year}{2010}).  
 
\bibitem{Nakaiprb}
\bibinfo{author}{Y. Nakai, T. Iye, S. Kitagawa, K. Ishida, S. Kasahara, T. Shibauchi, Y. Matsuda and T. Terashima}, 
\bibinfo{journal}{Phys. Rev. B.}
\textbf{\bibinfo{volume}{81}}, \bibinfo{pages}{020503(R)}
(\bibinfo{year}{2010}).  
 
\bibitem{Fukazawa}
\bibinfo{author}{H. Fukazawa, Y. Yamada, K. Kondo, T. Saito, Y. Kahori, K. Kuga, Y. Matsumoto, S. Nakatsuji, H. Kito, P. M. Shirage, K. Kihou, N. Takeshita, C.-H. Lee, A. Iyo, and H. Eisaki}, 
\bibinfo{journal}{J. Phys. Soc. Jpn.}
\textbf{\bibinfo{volume}{78}}, \bibinfo{pages}{083712}
(\bibinfo{year}{2009}). 

 \bibitem{Hashimoto2}
\bibinfo{author}{K. Hashimoto, A. Serafin, T. Tonegawa, R. Katsumata, R. Okazaki, H. Fukazawa, Y. Kohori, K. Kihou, C. H. Lee, A. Iyo, H. Eisaki, H. Ikeda, Y. Matsuda, A. Carrington, and T. Shibauchi}, 
\bibinfo{journal}{Phys. Rev. B.}
\textbf{\bibinfo{volume}{82}}, \bibinfo{pages}{014526}
(\bibinfo{year}{2010}).  

\bibitem{Miclea}
\bibinfo{author}{C. F. Miclea, M. Nicklas, H. S. Jeevan, D. Kasinathan, Z. Hossain, H. Rosner, P. Gegenwart, C. Geibel, F. Steglich}, 
\bibinfo{journal}{Phys. Rev. B}
\textbf{\bibinfo{volume}{79}}, \bibinfo{pages}{212509}
(\bibinfo{year}{2009}). 

\bibitem{Terashima}
\bibinfo{author}{T. Terashima, M. Kimata, H. Satsukawa, A. Harada, K. Hazama, S. Uji, H. S. Suzuki, T. Matsumoto, and K. Murata}, 
\bibinfo{journal}{J. Phys. Soc. Jpn.}
\textbf{\bibinfo{volume}{78}}, \bibinfo{pages}{083701}
(\bibinfo{year}{2009}). 

\bibitem{Uhoya}
\bibinfo{author}{W. Uhoya, G. Tsoi, Y. K. Vohra, M. A. McGuire, A. S. Sefat, B. C. Sales, D. Mandrus, S. T. Weir}, 
\bibinfo{journal}{J. Phys.: Condens. Matter}
\textbf{\bibinfo{volume}{22}}, \bibinfo{pages}{292202}
(\bibinfo{year}{2010}). 
 
\bibitem{Ren}
\bibinfo{author}{Z. Ren, Q. Tao, S. Jiang, C. Feng, C. Wang, J. Dai, G. Cao, Z. Xu}, 
\bibinfo{journal}{Phys. Rev. Lett.}
\textbf{\bibinfo{volume}{102}}, \bibinfo{pages}{137002}
(\bibinfo{year}{2009}).  
 
\bibitem{Sun}
\bibinfo{author}{L. Sun, J. Guo,  G. Chen, X. Chen, X. Dong, W. Lu, C. Zhang, Z. Jiang, B. Zou, Y. Huang, Q. Wu, X. Dai, Z. Zhao}, 
\bibinfo{journal}{Phys. Rev. B}
\textbf{\bibinfo{volume}{82}}, \bibinfo{pages}{134509}
(\bibinfo{year}{2010}).  

\bibitem{Matsubayashi}
\bibinfo{author}{K. Matsubayashi, K. Munakata, N. Katayama, M. Isobe, K. Ohgushi, Y. Ueda, N. Kawamura, M. Mizumaki, N. Ishimatsu, M. Hedo, I. Umehara, and Y. Uwatoko}, 
\bibinfo{journal}{arXiv:}
\bibinfo{pages}{1007.2889}
(\bibinfo{year}{2010}). 
 
\bibitem{Guo}
\bibinfo{author}{J. Guo, L. Sun, C. Zhang, G. Chen, J. He, X. Dong, W. Yi, Y. Li, X. Li, J. Liu, Z. Jiang, X. Wei, Y. Huang, Q. Wu, X. Dai, Z. Zhao}, 
\bibinfo{journal}{arXiv:}
\bibinfo{pages}{1008.2086}
(\bibinfo{year}{2010}).  
 
\bibitem{Kasahara10b}
\bibinfo{author}{S. Kasahara, K. Hashimoto, R. Okazaki, H. Shishido, M. Yamashita, K. Ikada, S. Tonegawa,
N. Nakata, Y. Sensyu, H. Takeya, K. Hirata, T. Shibauchi, T. Terashima and Y. Matsuda}, 
\bibinfo{journal}{Physica C}
(\bibinfo{year}{2010})
(\bibinfo{journal}{\textit{in press}}).
 
\bibitem{Alireza03}
\bibinfo{author}{P. L. Alireza and S. R. Julian}, 
\bibinfo{journal}{Rev. Sci. Instrum.}
\textbf{\bibinfo{volume}{74}}, \bibinfo{pages}{4728}
(\bibinfo{year}{2003}).  

\bibitem{Goh08}
\bibinfo{author}{S. K. Goh, P. L. Alireza, P. D. A. Mann, A.-M. Cumberlidge, C. Bergemann, M. Sutherland and Y. Maeno}, 
\bibinfo{journal}{Curr. Appl. Phys.}
\textbf{\bibinfo{volume}{8}}, \bibinfo{pages}{304}
(\bibinfo{year}{2008}).  

\bibitem{Osakabe}
\bibinfo{author}{T. Osakabe, K. Kakurai}, 
\bibinfo{journal}{Jpn. J. Appl. Phys.}
\textbf{\bibinfo{volume}{47}}, \bibinfo{pages}{6544}
(\bibinfo{year}{2008}).

\bibitem{Park}
\bibinfo{author}{J. T. Park, D. S. Inosov, C. Niedermayer, G. L. Sun, D. Haug, N. B. Christensen, R. Dinnebier, A. V. Boris, A. J. Drew, L. Schulz, U. Wolff, V. Neu, X. Yang, C. T. Lin, B. Keimer, V. Hinkov}, 
\bibinfo{journal}{Phys. Rev. Lett.}
\textbf{\bibinfo{volume}{102}}, \bibinfo{pages}{117006}
(\bibinfo{year}{2009}). 

\bibitem{jiang}
\bibinfo{author}{S. Jiang, H. Xing, G. Xuan, C. Wang, Z. Ren, C. Feng, J. Dai, Z. Xu, G. Cao}, 
\bibinfo{journal}{J. Phys.: Condens. Matter}
\textbf{\bibinfo{volume}{21}}, \bibinfo{pages}{382203}
(\bibinfo{year}{2009}). 

\bibitem{Onari}
\bibinfo{author}{S. Onari, H. Kontani}, 
\bibinfo{journal}{Phys. Rev. Lett.}
\textbf{\bibinfo{volume}{103}}, \bibinfo{pages}{177001}
(\bibinfo{year}{2009}).

\bibitem{Shishido}
\bibinfo{author}{H. Shishido, A. F. Bangura, A. I. Coldea, S. Tonegawa, K. Hashimoto, S. Kasahara, P. M. C. Rourke, H. Ikeda, T. Terashima, R. Settai, Y. Onuki, D. Vignolles, C. Proust, B. Vignolle, A. McCollam, Y. Matsuda, T. Shibauchi, and A. Carrington}, 
\bibinfo{journal}{Phys. Rev. Lett.}
\textbf{\bibinfo{volume}{104}}, \bibinfo{pages}{057008}
(\bibinfo{year}{2010}). 

\bibitem{Nakaiprl}
\bibinfo{author}{Y. Nakai, T. Iye, S. Kitagawa, K. Ishida, H. Ikeda, S. Kasahara, H. Shishido, T. Shibauchi, Y. Matsuda and T. Terashima}, 
\bibinfo{journal}{Phys. Rev. Lett.}
\textbf{\bibinfo{volume}{105}}, \bibinfo{pages}{107003}
(\bibinfo{year}{2010}). 

\bibitem{Duncan}
\bibinfo{author}{W. J. Duncan, O. P. Welzel, C. Harrison, X. F. Wang, X. H. Chen, F. M. Grosche, and P. G. Niklowitz}, 
\bibinfo{journal}{J. Phys.: Condens. Matter}
\textbf{\bibinfo{volume}{22}}, \bibinfo{pages}{052201}
(\bibinfo{year}{2010}). 

\bibitem{Dunstan}
\bibinfo{author}{D. J. Dunstan}, 
\bibinfo{journal}{Rev. Sci. Instrum.}
\textbf{\bibinfo{volume}{60}}, \bibinfo{pages}{3689}
(\bibinfo{year}{1989}). 

\bibitem{Hardy}
\bibinfo{author}{F. Hardy, P. Adelmann, T. Wolf, H. v. L$\mathrm{\ddot{o}}$hneysen, C. Meingast}, 
\bibinfo{journal}{Phys. Rev. Lett.}
\textbf{\bibinfo{volume}{102}}, \bibinfo{pages}{187004}
(\bibinfo{year}{2009}). 

\bibitem{Budko}
\bibinfo{author}{S. L. BudÕko, N. Ni, S. Nandi, G. M. Schmiedeshoff, and P. C. Canfield}, 
\bibinfo{journal}{Phys. Rev. B}
\textbf{\bibinfo{volume}{79}}, \bibinfo{pages}{054525}
(\bibinfo{year}{2009}). 



  





 
  

  
\end{thebibliography}
\end{document}